\documentclass[apj, twocolumn]{aastex631}
\usepackage{epstopdf}
\usepackage{graphicx}
\usepackage{wrapfig}
\usepackage[utf8]{inputenc}
\usepackage{amsmath}
\usepackage{natbib}
\usepackage{multirow}
\usepackage{longtable}
\bibpunct{(}{)}{;}{a}{}{,}
\usepackage{comment}
\usepackage{longtable}
\usepackage{color}
\definecolor{darkgreen}{RGB}{0,142,128}
\definecolor{darkblue}{RGB}{0,100,170}
\definecolor{darkpurple}{RGB}{150,0,150}

\usepackage{hyperref}
\hypersetup{colorlinks,citecolor=blue,linkcolor=blue}

\begin{document}
\title{An MHD Simulation of the Possible Modulations of Stellar CMEs Radio Observations by an Exoplanetary Magnetosphere}
%\title{Is there any influence of Exoplanetary atmosphere on the nature of the Radio burst? MHD simulations perspective}

%Exoplanets at Radio, UV and X-ray Wavelength
\author[0000-0002-7069-1711]{Soumitra Hazra}
\email{soumitra\_hazra@uml.edu, soumitra.hazra@gmail.com}
\affiliation{Lowell Center for Space Science and Technology, University of Massachusetts Lowell, 600 Suffolk Street, Lowell, MA 01854, USA}
\author[0000-0003-3721-0215]{Ofer Cohen}
\affiliation{Lowell Center for Space Science and Technology, University of Massachusetts Lowell, 600 Suffolk Street, Lowell, MA 01854, USA}
\email{ofer\_cohen@uml.edu}
\author[0000-0002-6118-0469]{Igor V. Sokolov}
\affiliation{Department of Climate and Space Sciences and Engineering, University of Michigan, 2455 Hayward St., Ann Arbor, MI 48109, USA}
\email{igorsok@umich.edu}

%\email{soumitra\_hazra@uml.edu, soumitra.hazra@gmail.com}
%\affiliation{Lowell Center for Space Science and Technology, University of Massachusetts Lowell, 600 Suffolk Street, Lowell, MA 01854, USA}

%\author[0000-0003-3721-0215]{Ofer Cohen}
%\affiliation{Lowell Center for Space Science and Technology, University of Massachusetts Lowell, 600 Suffolk Street, Lowell, MA 01854, USA}
%\email{ofer\_cohen@uml.edu}
%\author[0000-0002-6118-0469]{Igor V. Sokolov}
%\affiliation{Department of Climate and Space Sciences and Engineering, University of Michigan, 2455 Hayward St., Ann Arbor, MI 48109, USA}
%\email{igorsok@umich.edu}

\begin{abstract}
Type II radio bursts are the indicator of adverse space weather in a stellar system. These radio bursts are the
consequence of shock wave acceleration due to the coronal mass ejection (CME). In this study, we conduct a series of magnetohydrodynamic (MHD) simulations of CME-driven star–planet systems to investigate how close-in exoplanets modulate radio burst characteristics. We use a model for the stellar wind with a close-in
exoplanet, and a CME model based on the eruption of a flux rope. We are able to generate synthetic radio burst images from our MHD simulations. We find that radio burst like phenomena is most likely to be observed for moderately active solar like stars and close-in exoplanetary systems have significant influence on the nature of
radio burst spectrum. We find that when the exoplanet’s magnetic field is relatively weak, its magnetosphere compresses the CME plasma, increasing local density and shifting the radio emission to higher frequencies. Conversely, a strong planetary magnetic field results in a large magnetosphere that prevents effective CME-
shock development, producing weaker radio emission concentrated at lower frequencies, particularly at the flanks of the CME. For highly active solar-like stars, strong overlying stellar magnetic fields suppress the CME shock, greatly diminishing radio burst visibility. For HD 189733 (moderate stellar field), only intensity
difference is visible when the CME arrives the planet. We also do not find significant modulation in the radio emission by a close-in exoplanet system when the stellar magnetic field is complex. In summary, our findings highlight that the nature of the radio burst spectrum is strongly dependent on both the topology of the stellar
magnetic field and the magnetic strength of close-in exoplanets.

\end{abstract}


%\begin{keyword}
%% keywords here, in the form: keyword \sep keyword, up to a maximum of 6 keywords
%Magnetohydrodynamical Simulations \sep Exoplanets \sep Radio astronomy\sep Stellar Coronae 

%% PACS codes here, in the form: \PACS code \sep code

%% MSC codes here, in the form: \MSC code \sep code
%% or \MSC[2008] code \sep code (2000 is the default)

%\end{keyword}

%\end{frontmatter}
%\linenumbers
\section{Introduction}
Radio observations of stellar systems are one of the promising ways to study stellar magnetic activity and space weather of extra-solar planets \citep{Dulk1985, Callingham2024, Lazio2024}. Energetic explosive events in the solar-stellar system are often associated with sudden and intense radio emission, generally known as radio bursts \citep{Dulk1985}. Solar flares and coronal mass ejections (CMEs), two energetic explosive events in the solar system, release massive magnetic energy ($10^{33}$ ergs) in the form of particle acceleration and bulk plasma motion and create large-scale disturbances in the space weather of the solar system \citep[e.g.,][]{Hundhausen1997, Manchester2017}. Sufficiently strong Earth-directed CMEs have the potential to disturb the Earth's atmosphere, causing geomagnetic disturbances and damaging space-based instruments and vulnerable Earth-based infrastructures \citep{Kilpua2017}. The stellar counterpart of the solar CMEs, also known as stellar CMEs, may release much larger energies and play an essential role in shaping the atmosphere around that star \citep{Alvarado2018,2022SerAJ.205....1L, Alvarado22, Alvarado22a}. The enhanced particle and photon fluxes associated with stellar CMEs directly impact the exoplanet atmosphere and could drive atmospheric mass loss leading to atmospheric depletion \citep{Vidal2003, Lammer13a, Hazra2025a, Hazrag2025}. Steller CMEs and superflares are believed to be important in determining the planetary habitability \citep{Yamashiki2019, Hu2022}. However, despite their importance, direct observational evidence is almost non-existent for stellar CMEs \citep{Veronig2021, Namekata2022}. In this situation, the time-resolved radio dynamic spectrum (radio burst) offers one of the best indirect ways of detecting and characterizing coronal mass ejections from other stars.  A radio burst is a short period when stellar coronal radio emission is suddenly elevated above the background levels. Radio burst observations exist for the sun and a few other stars \citep{Villadsen2014, Feeney2021}. A deeper understanding of the radio burst mechanism is thus necessary to understand these extremely energetic events and their impact on the exoplanetary atmosphere \citep{Zarka2007, Klein2018}. This understanding is also important if we want to characterize the exoplanetary atmosphere and asses the condition for habitability. 

Two radio emission types, thermal and nonthermal, are associated with the CMEs. Most of the observed metric radio bursts associated with CMEs are nonthermal. The two most important types of non-thermal radio bursts for characterizing space weather are known as Type-II and Type-III radio bursts. Type II radio bursts are produced by the plasma emission from the electrons accelerated by shocks at the leading-edges of the outward propagating CMEs   \citep{Wild72, Robinson1985, Nelson85, Cairns11}. Type-II radio bursts generally last several minutes to hours and tend to slowly decrease in frequency as plasma density decreases when the shock wave propagates out of the solar corona. Type III radio bursts are very short lived (only few seconds) and associated with electron beams accelerated by reconnection events. One can see Figure 1 of \cite{Callingham2024} for a schematic of type-II and type-III radio bursts. On the other hand, thermal electrons can also generate detectable Bremsstrahulng radio emissions \citep{Sheridan1978}. Brighter nonthermal radio emissions generally mask thermal radio bursts, making them difficult to detect \citep{Bastian1997}. However, some previous studies indicate the possibility of detecting thermal radio emissions when their nonthermal counterpart is relatively weak \citep{Kundu1990, Gopalswamy1993}. In the case of the Sun, there are reports of direct CME detection via thermal bremsstrahlung radio emission \citep{Sheridan1978, Gopalswamy1992, Kathiravan2005, Ramesh2021}. Thermal radio emissions from other stellar systems are also detected \citep{Villadsen2014, Ogorman2017}. We also note that many CMEs are radio quiet \citep{Gopalswamy2008, Carley2020}.

Except for Type-II radio emissions, most nonthermal radio emissions are associated with the flaring mechanism. Type-II nonthermal radio emissions result from shock waves generated because of CMEs. Active stars flare frequently \citep[e.g.,][]{Kashyap2002}. Many observations of stellar flares in radio, UV, X-ray, and optical bands exist in the literature \citep{Macgregor2021, Jackman2021}. Superflare-like events are also observed in other stars, while these kinds of superflares are not observed in the Sun \citep{Candelaresi2014}. \cite{Shibata2013} argues that flux transport dynamo models can explain these kinds of stellar superflare events (see \cite{Hazra2014, Hazra2016} for the details of the flux transport dynamo model). \citep{Mullan2018} also investigated the impact of the flares on the exoplanetary atmosphere and habitability. On the other hand, it is difficult to observe stellar CMEs with our present instruments. \cite{Argiroffi2019} is the only study that reported the direct detection of stellar CMEs from blue-shifted cool coronal emission following a flare on the giant star HR9024. However, different proxies have been proposed for indirect evidence of stellar CMEs. Two promising proxies are: Doppler shifts in the UV wavelengths \citep{Vida2016} and X-ray continuous absorption \citep{Moschou2017}. \cite{Moschou2019} reviewed all investigated stellar CME candidates by these two methods.  Some studies also investigated the signal of filament/prominence eruptions and superflares as indirect evidence of stellar CMEs and successfully detected the H$\alpha$ line spectra of five superflares on EK Dra \citep{Namekata2022c, Namekata2022b, Namekata2024a, Namekata2024b}. Type II radio burst is another promising candidate for indirect detection of stellar CMEs; however, it has failed to do any positive detection \citep{Osten2015, Villadsen2017, Crosley2018}. 

\cite{Drake2013} suggest that CMEs mostly dominate strong winds of the active star. \cite{Cranmer2017} found that the mass loss rate of young stars of age less than 1GYr is mostly dominated by CMEs. Few previous studies found increased atmospheric mass loss by close orbit exoplanets because of the stellar CMEs \citep{Segura10, Drake2019, GHazra2022}. \cite{Alvarado2018} showed that for active M dwarf stars, the strong stellar magnetic field can partially or completely suppress the CMEs. This result indicates that stellar CMEs in active stars can be rare and their associated radio signal can be very weak compared to the solar case.   Some previous studies indicate the possibility of radio-quiet CME for M-dwarfs due to their strong surface magnetic field and corresponding large Alfv\'en speed \citep{Mullan2019, Villadsen2019}.  It is thus necessary to develop a good understanding of stellar CMEs, stellar radio emissions, and their interaction with the exoplanetary atmosphere.

In this study, we aim to investigate the impact of the exoplanetary magnetosphere and the stellar magnetic field on the nature of thermal radio emissions in the presence of CMEs. Here we first perform 3D MHD wind simulations to obtain a steady-state wind solution, and then we allow a CME to erupt from the stellar surface. Basically, we model the stellar corona without a nearby planet. Next, we place a planet near the star and investigate the impact of CMEs on the planetary magnetosphere. We perform this study by varying different parameters related to planetary and stellar magnetic fields.

We describe the details of our model in Section~\ref{Model}. We present the details of our results and discussion in Section~\ref{Results} and a Section~\ref{Discussion} respectively. Finally, we present our conclusion in the last section. 

\section{Model}
\label{Model}
\subsection{Star-Planet Interaction Model}
We use the Alfve\'en wave-driven solar atmosphere model (AWSoM) to obtain the steady-state stellar wind solution. In this model, we solve a set of nonideal magnetohydrodynamic (MHD) equations in conservative form, namely, mass continuity, momentum, magnetic induction, and energy equations. In addition, this model solves two additional equations to account for the propagation, reflection, and dissipation of the Alfv\'en wave energy—one for waves propagating parallel
to the magnetic field and another for waves antiparallel to the magnetic field (see \cite{vand14} for the details of this setup). As suggested by \cite{Sokolov2021}, the threaded-field-line model(TFLM) was used for modeling the transition region and chromosphere in a one dimensional manner, in order to accelerate the numerical convergence. 

We constrain the inner boundary conditions by the surface magnetic field data obtained from observed photospheric magnetogram data. We use different stellar surface magnetogram data for this study. Finally, we calculate the three-dimensional magnetic field using the potential field source surface extrapolation techniques \citep{Schatten1969} and initialize our stellar wind model. AWSoM model then solves all MHD equations to get the steady state solution for the stellar corona and wind. Different studies have extensively used this approach to obtain the steady-state stellar and solar wind solution \citep{vand14}. 

As our main aim here is to study the star-planet interaction, we model the planet via an additional boundary condition for the second body imposed in our computational domain. We refer the readers to \cite{Hazra2022} for a detailed description of the star-planet interaction setup. Please note that we do not consider any planetary orbital motion for this study. We fixed the planet's position along the line of sight direction of our telescope. We choose the planetary boundary number density and boundary temperature value as $10^7$ cm$^{-3}$ and $10^4$ K, respectively \citep{Cohen18}. 

\subsection{Flux Rope Model for the Coronal Mass Ejection}
We use the  Titov-D\'emoulin flux rope eruption model \citep{Titov1999} to model the coronal mass ejection (CME). In this flux rope eruption model, generally, we insert an unstable twisted flux rope in the background magnetic field obtained from our steady state wind simulation. Inserted flux rope then erupts due to the magnetic pressure/tension imbalance and generates coronal mass ejection. We study the evolution and propagation of the coronal mass ejection by a time-dependent simulations covering 120 minutes (real time). We chose our simulation time so that CMEs get sufficient time to reach the planetary magnetosphere position. This approach enables us to study the interaction of coronal mass ejection with the planetary magnetosphere in greater detail.  This type of numerical approach to model coronal mass ejection has been widely used in the existing literature of CME modeling. \citep{Manchester2008, Titov2014, Regnault2023}

Shape of the Titov-D\'emoulin flux rope is semicircular and its size is controlled by two geometrical parameters, namely, major radius and minor radius. Current of the TD-flux rope is concentrated at the center of the toroidal flux rope \citep{Titov2014}. Different parameters to describe an initial Titov-D\'emoulin (TD) flux rope are listed in Table 1. We choose the tilt angle and location of the flux rope in a way such that CME hits the planetary magnetosphere directly. For our study, we set all of these at $0$ degree. Height of the initial flux rope in our model is adjusted by burying the flux rope partially under the stellar surface, characterized by the parameter Depth (0.2 Solar Radius). Depth is basically the location of the toroid center below the photosphere level. Flux rope is loaded with mass and has an initial magnetic field. Magnetic field at the center of the toroidal flux rope, denoted as $B_c$, is generated by the current inside the toroidal flux rope. We can calculate the free energy of the CMEs using the toroidal current and the geometric properties of the flux rope. In summary, one can control the free energy of the CMEs by modifying the magnetic field at the center and geometric properties of the flux rope. In our model seup, we initialize the eruption with a magnetic free energy $E_B^{FR} \approx 2 \times 10^{33}$ erg (associated with the magnetic field at the center ($B_c$) 10 Gauss and corresponding toroidal current $5 \times 10^{14}$ Ampere) and loaded the flux ropes with an initial mass of $4.0 \times 10^{14}$ gm. Our choice of loaded mass ($M^{FR}$) and free energy ($E_B^{FR}$) is close to the high-end values observed in solar CMEs
\citep{Gopalswamy2009, Toriumi2017}. One can see \cite{Roussev2003} and \cite{Sokolov2023} for a detailed discussion of TD flux rope and its numerical implementation in the SWMF framework.\\

\begin{table}[]
   \caption{Initialization parameters of 
the TD flux-rope \label{tab_1}}
\smallskip
    \centering
    \begin{tabular}{c c c}
    \hline \hline
    Parameter  ~~~~ ~~~&   Value ~~~~~~~~& Unit\\
    \hline
     Tilt angle$^{\dagger}$ & 0.0 & deg\\ 
     Longitude & 0.0 & deg\\
Latitude & 0.0 & deg\\
Major Radius & 0.5 & R$_s$\\
Minor Radius & 0.2 & R$_s$\\
Depth & 0.2 & R$_s$\\
Mass ($M^{\rm FR}$) & $4 \times 10^{14}$ & gm\\
Magnetic field at the \\
Center ($B_c$) & 10 & Gauss \\
\smallskip\\
 \end{tabular}
 \tablenotetext{}{\hspace{-0.1cm}$^{\dagger }$Measured with respect to the stellar equator in the counter clock-wise direction.}
\end{table}

\subsection{Synthetic Images of Stellar Corona in Radio and Radio Burst Mechanism}
The AWSoM model has a feature to generate synthetic radio images of the stellar corona from \cite{Moschou18}. The method considers the radio emission due to the thermal bremsstrahlung coronal emission. We trace the propagation of radio waves through the circumstellar medium of the nonuniform density by the ray-tracing algorithm developed by \cite{Benkevitch10, Benkevitch12}. This approach also accounts for the effect of radio wave refraction, as radio wave suffers more refraction due to their strongly varying refraction index. We refer the readers to \cite{Moschou18} and \cite{Hazra2022} for the complete description of the radio images tool. 

The radio burst mechanism represents the sudden increase of coronal radio emission for a short period. These events are generally associated with strong disturbances in the coronal atmosphere, and the increase in radio emission may depend on various factors, namely, plasma parameters of the source region, amount of released energy, properties of the coronal atmosphere, etc. Using synthetic radio images of different radio frequencies obtained from our model, we generate the dynamic radio spectrum (radio frequency vs. time plot) by the method of interpolation. This enables to compare these simulated dynamic radio spectrums against the observed dynamic radio spectrum obtained from radiospectrograph. 

In our modeled, thermally-driven radio burst, the dominant effect is the formation of a shock in front of the CME. The shock compresses and heats the surrounding plasma during its propagation through interplanetary medium. In general, higher compression (stronger shock) is responsible for the generation of higher radio frequency, while weaker compression (weaker shock) is responsible for the generation of lower radio frequencies. CME generated shocks are stronger close to the surface and they become weaker during their propagation through the interplanetary space. High- to low-frequency drift of the dynamic radio spectrum is due to the fact that the shock propagates outward, expands, and its compression gets weaker. Thus, higher radio frequencies appear first and the the frequency drifts to lower values. This is why Type II radio bursts are useful to estimate the e.g., CME speed. 

Here, we aim to investigate the impact of the close-in exoplanet on the radio burst in three possible ways: i) the frequency range (possible emission at higher frequencies); ii) a time shift in the radio burst evolution; and iii) a change in the slope of the frequency drift. In this study, we assume that stellar radio flux is actually observable from earth.  As we are only interested to study the modulation in radio intensity due to CME propagation and our work is only relevant for stars with observable radio flux from earth, we do not provide the actual magnitude of radio flux in units of Jy.

\section{Results}
\label{Results}
In this section, we aim to characterize radio burst signals from the computational modeling point of view. We first perform a steady-state for the ambient stellar wind. Next, we use that steady-state solution as an initial condition and perform the time-dependent simulation where we introduce the CME. As the CME propagates into space with high velocity (faster than the local Alfv\'en speed), it is shocked at its front, creating a compressed CME front. In reality, electrons are accelerated at the compressed CME shock, radiating synchrotron emission, which is visible as a Type II radio burst. Even though we cannot capture this particle acceleration in the MHD model, the compressed shock in our simulation will be visible as an increase radio emission in our synthetic images. Thus, we can still relate the synthesized radio emission to the physical evolution of the shock, and the possible radio manifestation of the planetary impact on the shock evolution, which is the goal of this paper.

\subsection{Star with Simple Dipolar Magnetic Field}
We first consider the case of a star with a simple solar-like dipolar magnetic field of 10G. Our choice of other stellar parameters is similar to the HD189733 system: $R_*=0.75R_\odot$, $M_*=0.5M_\odot$, and $P_{rot}=4.85$ days. We place the planet at a distance of $10~R_*$ along the $x=0$ axis. The synthetic radio telescope line-of-sight (LOS) is also along the $x=0$ line, so that the planet is exactly between the observer and the star. As we do not know the planetary magnetic field strength, we formulate three specific cases with different planetary magnetic fields for our study. We chose the planetary magnetic field of $0.3$ G (Earth-like), 1 G (a moderate field), and 6 G (very high planetary magnetic field) to design our study. For reference, Jupiter's field strength is of the order of $4$ G. This simplified setting allows us to qualitatively study the radio burst phenomena regardless of the particular star and planet we use, and to isolate the main effects without accounting for additional complexity that is not related to the impact of the planet on the radio burst signature. 

\begin{figure*}[!ht]
\centering
%\begin{tabular}{cc}
\includegraphics*[width=1.0\linewidth]{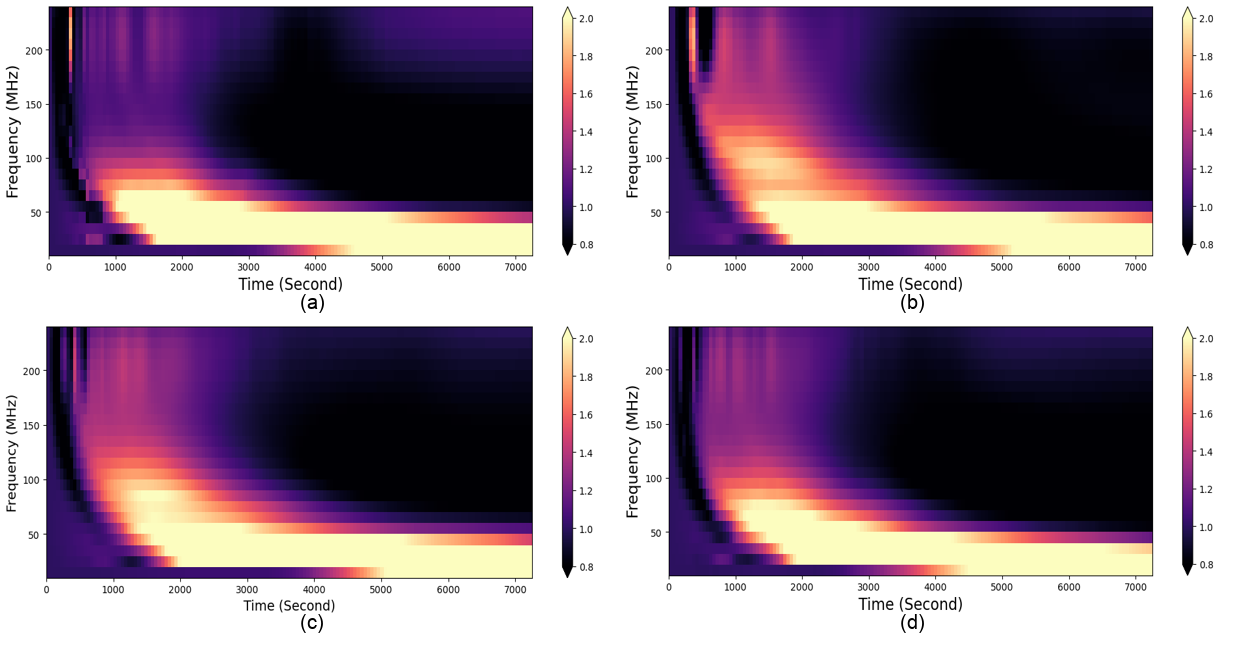} \\
\caption{\footnotesize{Figure (a) shows the simulated dynamic radio spectrum when there is no planet near the star. Figure (b), (c) and (d) corresponds to the same but with planet having different planetary magnetic field strengths, namely, 1 G, 3G, and 6 G respectively. In this case, our star has a dipolar magnetic configuration. The planet is placed at 10 stellar radii apart from the central star.}}
\label{fig:radio-dipole}
\end{figure*}

Figure~\ref{fig:radio-dipole}(a) shows the simulated dynamic radio spectrums obtained from our simulations when there is no planet nearby the central star. Figure~\ref{fig:radio-dipole}(b), (c), and (d) show the radio burst spectrum when a planet nearby the central star for different planetary magnetic field strengths, namely, 1G, 3G and 6G respectively. In all cases, we find that radio burst drifts slowly from high to low frequencies over the course of a few tens of minutes. Although our modeled dynamic radio spectrum is thermal in nature, it looks quite similar to what we observed in the case of the Sun. Figure~\ref{fig:radio-dipole}(b), (c), and (d) also indicate that high radio frequencies are more likely to be generated when there is a planet nearby the central star. We note that high to low-frequency drift speed is around 0.1-0.5 MHz per second when there is no planet near the central star (see Figure~\ref{fig:radio-dipole}a). High to low-frequency drift speeds are little bit higher (around 0.5-0.7 MHz per second) when there is a planet near the central star. Radio bursts are more likely to be visible in higher radio frequencies when there is a nearby planet.

\begin{figure*}[!ht]
\centering
%\begin{tabular}{cc}
\includegraphics*[width=0.9\linewidth]{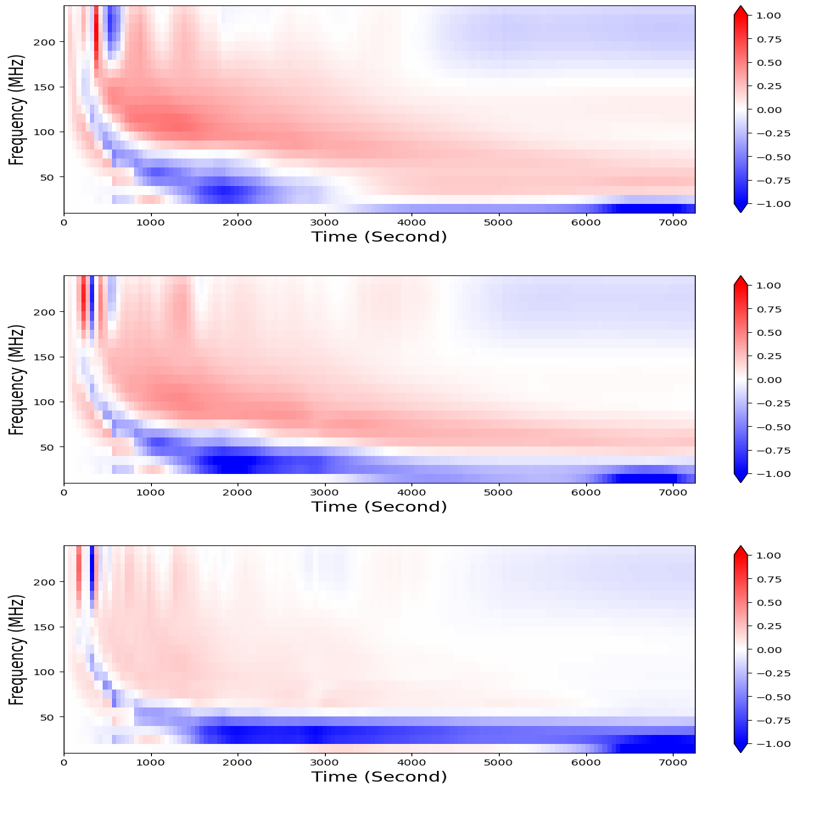} \\
\caption{\footnotesize{Figure shows the modulation of radio emission in presence of planet. Top panel shows the modulation when planet with magnetic field strength of 1 Gauss is present near the central star. Middle and bottom panel shows the same but planets with magnetic field strength of 3 and 6 Gausses respectively. In this case, our star has a dipolar magnetic configuration. The planet is placed at 10 stellar radii apart from the central star.}}
\label{fig:radio-dipole-diff}
\end{figure*}

Figure~\ref{fig:radio-dipole-diff} shows how the planetary presence modulates the observed radio emission signal. Top panel shows the modulation of radio emission when the planetary magnetic field strength is weaker (only 1 Gauss). While middle and bottom panel shows the same when the planetary magnetic field is relatively higher (3 Gauss and 6 Gauss respectively). We take the difference of radio emissions with planetary presence and without planetary presence to calculate this modulation. Comparison between these three plots suggest two things - first, if there is a close-in exoplanet near the star, higher radio frequencies are more likely to be produced; second, higher radio frequencies near the star is generated much when magnetic field of the nearby planet is comparatively weaker. First indication suggests that presence of exoplanets near the star helps to develop stronger shocks when CME propagates through the atmosphere region between the star and the close-in exoplanet. However, second indication suggests something different. We already discussed that stronger shocks are responsible for the generation of higher radio frequencies and we also know that higher planetary magnetic field creates larger planetary magnetosphere. Larger planetary magnetosphere does not leave much room for the evolution of CME shock making the CME shock weaker. In summary, if the planetary magnetic field is high, CME generated shocks will be weaker and not be able to generate higher radio frequencies for significant time. We note that planet can also be a source of radio signal as star-planet interaction can produce radio signal. However, previous studies suggest that CME driven shock would completely shut down the planet-induced stellar radio emission via electron-cyclotron maser instability \citep{Saur2013, Kavangh2021}. 

\begin{figure*}[!ht]
\centering
%\begin{tabular}{cc}
\includegraphics*[width=0.90\linewidth]{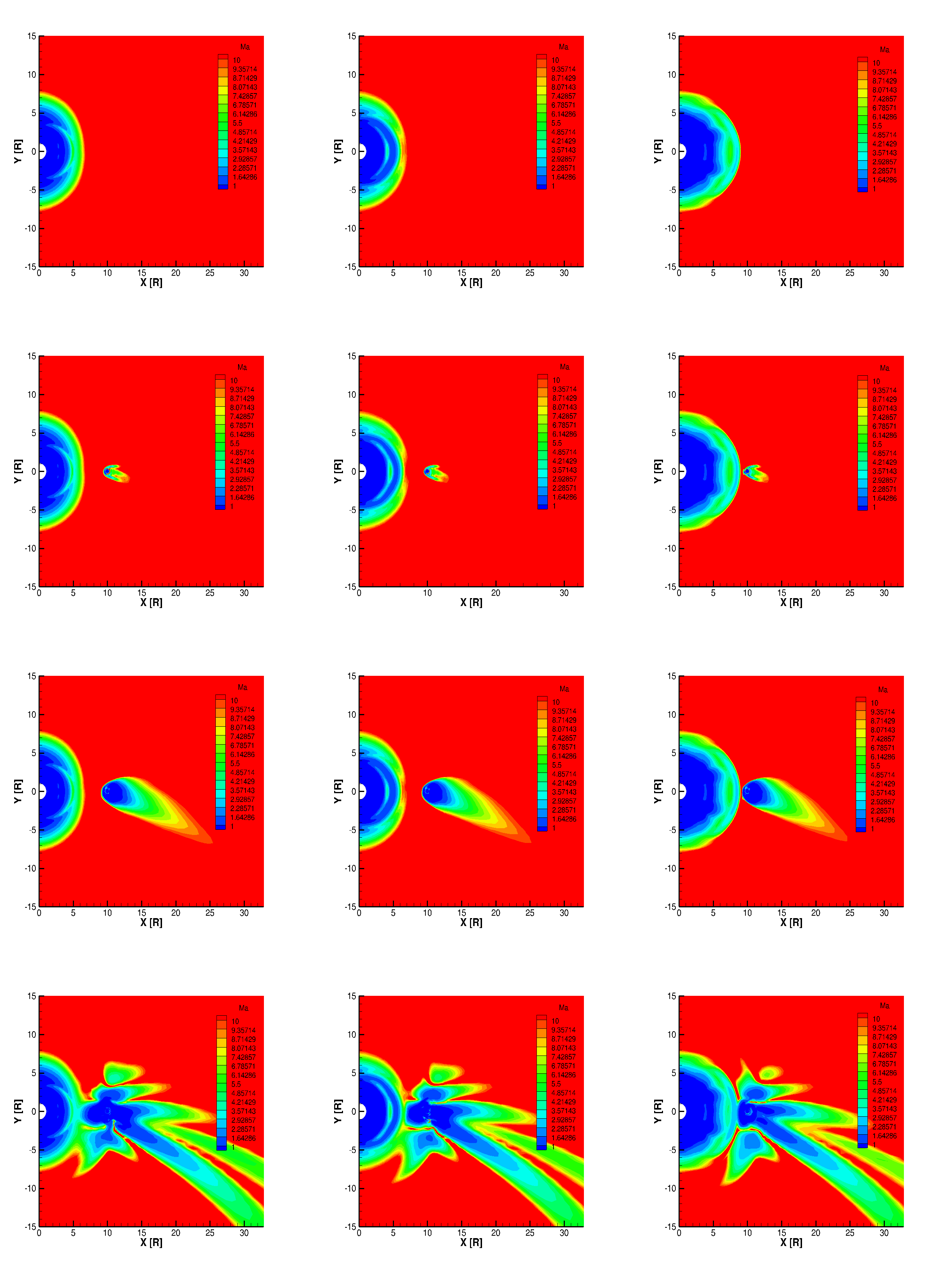} \\
\caption{\footnotesize{Figure shows the evolution of velocity in X-Y plane. First, second and third columns shows the evolution of velocity in X-Y plane after 30 minutes, 1 hours and 2 hours of CME eruption. While first, second, third and fourth row corresponds to the cases of no planet, planet with 1 Gauss planetary field, planet with 3 Gauss planetary field, and planet with 6 Gauss planetary fields respectively. While . In this case, our star has a dipolar magnetic configuration and the planet is placed at 10 stellar radii apart from the central star. Here, we represent the velocity in terms of Mach number.}}
\label{fig:evol-vel}
\end{figure*}

Figure~\ref{fig:evol-vel} shows the velocity evolution in the X-Y plane after 30 minutes, 1 hours and 2 hours of the CME eruption. Different rows of the Figure~\ref{fig:evol-vel} shows different case scenarios, namely, no planet, planet with 1 G, 3 G and 6G magnetic fields respectively. In the plot, we represent the background velocity in terms of the Mach number. Top panel (row) of Figure~\ref{fig:evol-vel} shows the scenario when there is no planet near the central star and we notice that shock associated with the CME moving slowly when we go outwards the central star. However, the presence of a planet near the central star impacts the shock wave propagation significantly (see two middle and bottom panels of Figure~\ref{fig:evol-vel}). The speed of the shock is actually reduced by the planetary magnetosphere, basically, CME slows down when it interacts with the planetary magnetosphere. Our results also indicate that the slowing down of CME depends on the strength of the planetary magnetosphere. In the case of 1G and 3G planetary magnetic field strength, the magnetosphere pushes against the CMEs, making the shock weaker (see two middle panels of Figure~\ref{fig:evol-vel}). Weakening of the shock is more in the case of 3G planetary magnetic field scenario compared to the 1G planetary magnetic field scenario. However, in the case of 6G planetary magnetic field strength, CME is completely disrupted and shock is weakened significantly (see bottom panel of Figure~\ref{fig:evol-vel}).

In one of our earlier study (see \cite{Hazra2022}), we have shown that the size of the planetary magnetospheres is bigger for higher planetary magnetic fields. Magnetosphere size in the 6G planetary magnetic field scenario is bigger compared to 1G and 3G planetary magnetic field scenarios. As the planetary magnetosphere size is bigger for the higher planetary magnetic field, the interaction of the planetary magnetosphere with the CME happens much earlier in the higher planetary magnetic field scenario compared to the lower planetary magnetic field scenario. One can thus expect a higher weakening of shocks or slowing down of CMEs in the higher planetary magnetic field scenario, our result also indicates the same. It is now well known that the nature of the dynamic radio burst spectrum depends on the nature of the shock wave. It is thus expected that the presence of the planet in a stellar system will have an impact on the stellar radio burst spectrum. Our results also suggest that higher radio emissions are much more likely when there is a planet near the star. Some part of the energy of the accelerated or decelerated plasma is most likely radiated at radio frequencies. It is also possible that the planetary magnetosphere here acts as a source of radio emission as suggested earlier \citep{Lazio18, Lynch18}. The planetary magnetosphere probably extracts energy from the stellar wind and CME-associated shocks and radiates some part of this energy, likely at radio frequencies.

Type II radio bursts are suggested as a promising candidate for the indirect detection of stellar coronal mass ejection \citep{Osten2015, Moschou2019}. One can probe the CME internal magnetic field during their propagation through the stellar atmosphere by measuring the linear polarization of radio signals \citep{Mancuso2000, Jackson2020, Jackson2023}. Linearly polarized radio emissions when passes through the magnetized plasma is subject to a rotation of its plane of polarization, known as Faraday Rotation. As the degree of rotation depends on the density and magnetic field along the path, thus one can probe the CME internal magnetic field by measuring the Faraday Rotation \citep{Mancuso2000, Kooi2017}. Observing stellar CME and associated radio bursts are thus utmost necessity. Although no observation of a stellar radio burst exists at the present moment, our simulated results clearly suggest that it is possible to observe radio burst signals for a moderately active solar-like star. Our result also indicates that stellar systems with the presence of planets are much more likely to be observed at higher radio frequencies.

\subsection{Star with Very Strong Dipolar Magnetic Field: Is it possible to observe the radio burst for highly active solar-like stars?}
Our results already indicate the possibility of observing stellar radio bursts for moderately active stars. However, the question remains, is it possible to observe radio bursts for highly active solar-like stars? To investigate this issue, we consider a solar-like star with a very strong 100G dipolar magnetic field. We formulate two specific cases for this investigation - in the first scenario, we assume that there is no planet in the stellar system, and in the second scenario, we assume the presence of a planet near the central star. In the second scenario, we place a hot Jupiter-like planet at a 10 stellar radii distance apart from the central star.

\begin{figure*}[!ht]
\centering
%\begin{tabular}{cc}
\includegraphics*[width=0.9\linewidth]{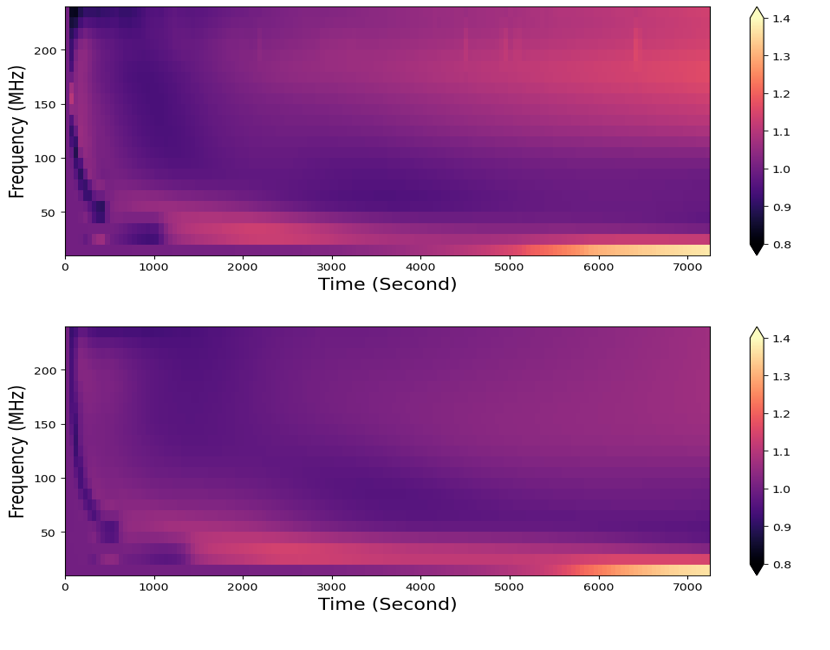} \\
\caption{\footnotesize{Top panel shows the simulated dynamic radio spectrum when there is no planet near the star. The bottom panel shows the same but with the planet having a planetary magnetic field strength of 3 Gauss. In this case, our star has a strong dipolar magnetic configuration having a stellar magnetic field of 100G. The planet is placed at 10 stellar radii apart from the central star.}}
\label{fig:high-stellar-field}

\includegraphics*[width=0.85\linewidth]{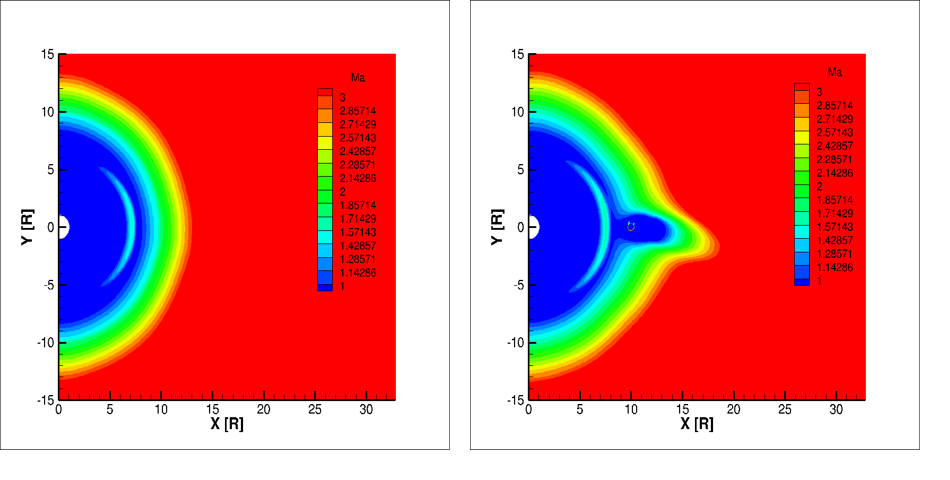} \\
\caption{\footnotesize{Left panel shows the evolution of velocity after one hour of CME eruption when there is no planet near the star. The right panel shows the same but with the planet having a planetary magnetic field strength of 3 Gauss. In this case, our star has a strong dipolar magnetic configuration having a stellar magnetic field of 100G. The planet is placed at 10 stellar radii apart from the central star. We represent the background velocity in terms of Mach number.}}
\label{fig:high-stellar-field-mach}
\end{figure*}

Figure~\ref{fig:high-stellar-field} shows the dynamic radio spectrums obtained from our simulations of highly active solar-like stars. The top panel shows the dynamic radio spectrum when there is no planet near the star and the bottom panel shows the same when there is a planet near the star. In both scenarios, we do not find any strong signature of a radio burst. A radio burst is basically a sudden increase in the radio emission. Radio emission fluxes are not very high in both scenarios. Our results suggest that radio burst is very unlikely to be observed for highly active solar-like stars. Figure~\ref{fig:high-stellar-field-mach} shows the velocity evolution after one hours of the CME eruption. Left panel corresponds to the situation when there is no planet near the central star, while right panel corresponds to the situation when there is a planet near the central star. Comparison between two panels clearly shows that propagation of shock waves are impacted by the close-in exoplanets. However, second panel also indicates that CMEs are suppressed by strong overlying fields. Basically, CME related shocks are not able to propagate outwards quickly due to the strong overlying magnetic fields and become weaker very quickly.

Type II radio bursts are the evidence for the presence of shocks in the stellar atmosphere. These shocks generally associated with the coronal mass ejection like events and can emit radio signals from submeter to kilometer wavelengths. Stronger shocks results in higher radio frequency generation in the stellar system. Our present simulation for highly active stars indicates the confinement of CME associated shocks by large scale overlying magnetic fields present in the stellar system. Shocks are weakened quickly and resulting in a very weak radio signal. Previous studies also indicate the suppression of CME like events by large scale magnetic fields and consequent brightening of solar corona in X-ray (see \citep{Alvarado2019}). Our result suggest that confinement of CME like events by strong overlying magnetic fields will lead to non-observance of radio burst like events. However, we note that our chosen CME energy range is solar-like, around $10^{33}$-$10^{34}$ ergs. It is possible that in the case of highly active stars, CME energies could be significantly higher. In such scenarios, these stellar super-CMEs might overcome magnetic confinement and generate strong radio emissions. However, there is currently no observational evidence for the existence of such super-CMEs.

\subsection{HD189733 Star}
All of our results till now based on the stellar system having ideal dipolar magnetic field structure. However, in reality, magnetic field structure of the stars are highly complex. Topology of the stellar surface magnetic field significantly influence the coronal and solar wind structure (see \cite{Hazra2021}). Motivated by these facts, here we choose a realistic stellar system HD189733 to investigate this issue further. This stellar system is also assumed to have a close-in exoplanet.

\begin{figure*}[!ht]
\centering
%\begin{tabular}{cc}
\includegraphics*[width=0.9\linewidth]{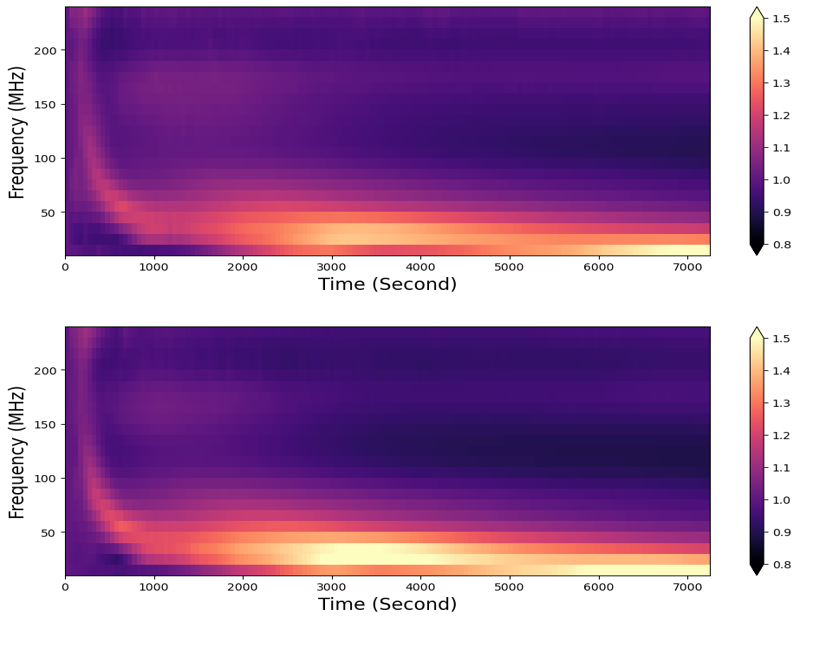} \\
\caption{\footnotesize{Top panel shows the simulated dynamic radio spectrum when there is no planet near the central star. The bottom panel shows the same but with the planet having a planetary magnetic field strength of 3 Gauss. Here, our selected star is HD189733 and the planet is placed at 10 stellar radii apart from the central star.}}
\label{fig:hd189733}
\includegraphics*[width=0.85\linewidth]{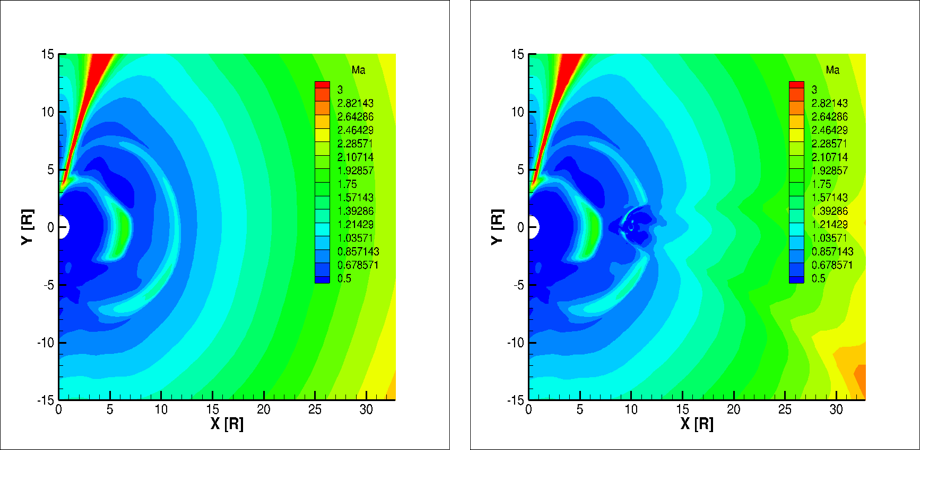} \\
\caption{\footnotesize{Left panel shows the evolution of velocity after the one hour of CME eruption when there is no planet near the central star. The right panel shows the same but with the planet having a planetary magnetic field strength of 3 Gauss. Here, our selected star is HD189733 and the planet is placed at 10 stellar radii apart from the central star. We represent the background velocity in terms of Mach number.}}
\label{fig:hd189733-mach}

\end{figure*}

Top panel of Figure~\ref{fig:hd189733} shows the dynamic radio spectrums obtained from our simulations for the star HD189733 when there is no planet near the star. The bottom panel shows the same but one planet is present near the central star. Both situations show a similar kind of dynamic radio spectrum. Basically, planetary presence does not have much impact on the radio burst spectrum except the increase of the power of low frquency radio emission when there is a nearby planet. However, we think this result is probably due to the fact that the magnetic field structure of the HD 189733 star is much more complex compared to our previous simple dipolar structure. Average magnetic field strength of the HD 189733 stellar system is also quite higher (around 40 Gauss) compared to our simple dipolar magnetic field structure. Previous studies also indicate that the topology of the surface magnetic field controls the coronal structure \citep{Hazra2022}. Please note that we also do not consider the presence of active regions in our study. It is now well known that the presence of active regions can significantly increase the coronal radio emission.

Left panel of Figure~\ref{fig:hd189733-mach} shows the velocity evolution after one hours of the CME eruption when there is no planet near the central star. Right panel shows the same but there is now a planet near the central star. Comparison between two plot indicates that although planetary magnetosphere modifies the shock structure but the modification is not very significant. That's why the dynamic radio burst spectrum looks almost similar in both cases. Probably presence of planet with very higher magnetic field is necessary to disturb the shock structure significantly.

%\begin{figure*}[!ht]
%\centering
%\begin{tabular}{cc}
%\includegraphics*[width=1.0\linewidth]{Figure5_paper.png} \\
%\caption{\footnotesize{Top panel shows the simulated dynamic radio spectrum when there is no planet near the central star. The bottom panel shows the same but with the planet having a planetary magnetic field strength of 3 Gauss. Here, our selected star is the Sun and the planet is placed at 10 stellar radii apart from the central star.}}
%\label{fig:sun}
%\end{figure*}

\section{Summary and Discussion}
\label{Discussion}
{\it Estimating the Space Weather of a Stellar System:}\\
It is necessary to understand the nature of stellar coronal mass ejections and their impact on the nearest planetary atmosphere for a better understanding of the space weather of that stellar system. Better understanding of the CME driven shocks will lead us to estimate the effectiveness of the coronal mass ejection. Radio observation is the only way to understand the strength of the shocks in a stellar system. Faster and brighter radio bursts is an indicator of the  larger shock strengths and thus more likely to have an larger impact on the planetary system. Radio observations also provide us an alternative indirect way to determine the coronal magnetic field via Faraday rotation measurement \citep{Mancuso2000}.

Our MHD simulation result suggests that it is possible to observe radio burst phenomena for moderately active solar like stars. These radio burst events will give us information regarding the shocks present in the stellar atmosphere. In summary, observed radio bursts will help us to estimate space weather of that stellar system.
\\

{\it The Potential of Radio Burst Observations and Simulations:}\\
Stellar type II radio burst is one of the best means for identifying and characterizing stellar CME or super-flare. Observing type II radio burst is also the only window to understand the particle acceleration process in the stellar atmosphere due to coronal mass ejection or superflare. However, previous attempts for observing type II radio bursts for active stars with the Very Large Array (VLA) have been failed \citep{Crosley2018a, Crosley2018, Crosley2018c, Villadsen2019}. Scintillation of background radio sources offers an alternative way for observing stellar CMEs. In case of solar CMEs, \cite{Manoharan2010} measured the scintillation for a large number of radio sourses and able to reconstruct three dimensional views of the propagationg coronal mass ejections. However, till now, there is no report of the scintillation measurement for other active stars.  

In our present study, we focuses on the potential of radio burst observations from the MHD modeling point of view. Our results indicate the possibility of observing radio burst spectrum for solar-like active stars. Our simulated radio burst signals for solar-like active stars are very strong.  In case of highly active stars, we do not find strong signature of radio burst as strong overlying magnetic field lines suppress the acceleration of evacuated coronal materials due to CMEs. However, we note that energy of CMEs in our model is solar-like. Highly energetic super-CMEs may break the barrier of strong overlying magnetic field and produce strong radio signals.  The near-future Square Kilometre Array may have the potential of detecting stellar CMEs in nearby solar like active stars.

We note that beyond radio observations, there are other observational ways to detect stellar CMEs or superflares \citep{Garcia2023}. Chromospheric and coronal emission lines supposed to show blueshifted doppler emission components during CMEs as it carries away the coronal materials. Few stellar CMEs have been recently observed by this method (see \cite{Moschou2019, Namekata2022} for a detailed discussion). Another potential way to observe stellar CMEs is the EUV dimming. As CMEs carries away the coronal material, it is expected that there will be a dimming in the EUV wavelength observation. This type of EUV dimming due to coronal mass ejection is already observed in the case of the Sun. \cite{Notsu16} found superflares in the active young M-dwarf stars and binary stars by analysing the Kepler photometric data. One may see the PhD thesis of the Yuta Notsu for a detailed discussion of superflare detection in active stars \citep{Notsu2019}. In summary, we need multiwavelength observations (in optical, EUV, radio and X-ray) to observe and characterize the stellar superflares and CMEs \citep{Namekata2024a, Namekata2024b}.\\

{\it Impact of the Exoplanet presence on the Radio Burst spectrum:}\\
\cite{Rubenstein2000} suggest that solar like stellar system with a hot jupiter exoplanet is an ideal candidate for a stellar system having superflares. Previous studies also indicate that we never observe superflares in our solar system as our solar system does not have any hot jupiter planets \citep{Schaefer2000}. That means presence of hot jupiter exoplanet in a stellar system is a necessary condition for having superflares. However, some other studies suggest that presence of hot jupiter exoplanet is not necessary for the generation of superflares in a stellar system \citep{Shibata2013, Notsu16}. We believe that even if presence of hot jupiter exoplanet is not necessary for the generation of superflares or highly energetic CMEs, the presence of hot jupiter like exoplanet may have significant impact on the CME/superflare associated shock wave propagation mechanism. Radio burst mechanism is the observational signature of the shock wave propagation mechanism in the stellar atmosphere.

In our study, we investigated the impact of close orbit hot jupiter exoplanet on the radio burst spectrum and find that stellar systems with close-in hot jupiter exoplanets are much more likely to be observed at higher radio frequencies. However, we note that this result is highly dependent on stellar magnetic field properties. We do not find any significant impact of close-orbit hot jupiter exoplanet on the radio burst spectrum when stellar surface magnetic field topology is relatively complex. Previous studies also indicate the importance of stellar/solar surface magnetic field properties in determining the coronal magnetic field properties \citep{Hazra2015, Hazra2021}

\section{Conclusion}
\label{Conclusion}
Observing and characterizing stellar coronal mass ejections and
their interactions with the exoplanetary atmosphere is crucial for the understanding of stellar systems and their space weather. In this study, we perform magnetohydrodynamic (MHD) simulations of CME-driven star-planet interactions to investigate the nature of the resulting radio burst spectrum for stellar systems. Without direct observation of stellar CMEs, dynamic radio burst spectrums often considered as an indirect proxy of stellar CMEs. In our study, we choose different surface magnetic field topologies for our central star. We sometimes place one hot Jupiter planet near our central star to investigate the interaction of stellar CMEs with the close-in exoplanets. Our simulations aim to assess the feasibility of detecting stellar CMEs and their planetary interactions via radio wavelength observations.

Our simulations demonstrate the feasibility of observing radio burst spectrums for moderately-active solar-like stars. In our simplistic, idealized CME-driven stellar wind simulation scenario (when the stellar surface magnetic field is dipolar), we are able to generate synthetic radio burst spectrums. Our model-generated radio burst spectrums are type II in nature as they show the trends of a slow decrease in radio frequency when the CME-generated shock wave moves out of the solar corona. When we place a close-orbit hot Jupiter exoplanet near the central star, we find significant modulation in the radio-burst spectrum due to the close-in exoplanet. However, the nature of impact (modulation) depends on the magnetic field strength of close-in exoplanet and stellar surface magnetic field topology.

In the case of our simulations with highly active solar-like stars
(the stellar surface magnetic field is very high), we do not find any clear signature of a radio burst. CME in our model setup is completely suppressed by strong-overlying stellar magnetic fields, making radio signals much weaker compared to the solar scenario. However, we note that the energy of the CMEs in our setup is solar-like. It is possible that super-CME events, with much higher energies, could overcome this magnetic suppression and produce detectable radio signatures–though such events have yet to be observed.

Additionally, we note that our study is based on the comparatively simpler stellar surface magnetic field topology. Highly complex coronal magnetic field structures (containing many coronal hole/open magnetic field regions) may have an impact on the nature of the radio burst spectrum. Our model also considers only thermal radio emission, not coherent (nonthermal) radio emission. Coordinated multiwavelength observations–spanning optical, radio, EUV, and X-ray–are essential for the detection and characterization of stellar CMEs. MHD simulations, such as those presented here, can provide valuable constraints and interpretation for such observations.

\section*{Acknowledgements}
This work is supported by NASA grant 80NSSC20K0840. Simulation results were obtained using the (open source) Space Weather Modeling Framework, developed by the Center for Space Environment Modeling, at the University of Michigan with funding support from NASA ESS, NASA ESTO-CT, NSF KDI, and DoD MURI. The simulations were performed on NASA's Pleiades cluster under SMD-20-52848317.

\bibliographystyle{yahapj} 
\bibliography{reference_radioburst}

\end{document}